\documentclass[twocolumn]{aastex61}

\newcommand\aastex{AAS\TeX}

\received{\today}
\revised{\***, 2018}
\accepted{\***,2018}
\submitjournal{ApJ}

\shorttitle{\aastex\ ALMA observations of G9.62+0.19}
\shortauthors{Liu et al.}

\shorttitle{Compressed magnetic field in the magnetically-regulated global collapsing clump of G9.62+0.19} \shortauthors{Liu et al.}

\begin{document}

\title{Compressed magnetic field in the magnetically-regulated global collapsing clump of G9.62+0.19}
\correspondingauthor{Tie Liu}
\email{liutiepku@gmail.com}

\author{Tie Liu}
\affiliation{Korea Astronomy and Space Science Institute 776, Daedeokdae-ro, Yuseong-gu, Daejeon, 34055, Republic of Korea}
\affiliation{East Asian Observatory, 660 N. A'ohoku Place, Hilo, HI 96720, USA}

\author{Kee-Tae Kim}
\affiliation{Korea Astronomy and Space Science Institute 776, Daedeokdae-ro, Yuseong-gu, Daejeon, 34055, Republic of Korea}

\author{Sheng-Yuan Liu}
\affiliation{Academia Sinica, Institute of Astronomy and Astrophysics, P.O. Box 23-141, Taipei 106, Taiwan}

\author{Mika Juvela}
\affiliation{Department of Physics, P.O.Box 64, FI-00014, University of Helsinki, Finland}

\author{Qizhou Zhang}
\affiliation{Harvard-Smithsonian Center for Astrophysics, 60 Garden Street, Cambridge, MA 02138, USA}

\author{Yuefang Wu}
\affiliation{Department of Astronomy, Peking University, Beijing 100871, China}

\author{Pak Shing Li}
\affiliation{Astronomy Department, University of California, Berkeley, CA 94720}

\author{Harriet Parsons}
\affiliation{East Asian Observatory, 660 N. A'ohoku Place, Hilo, HI 96720, USA}

\author{Archana Soam}
\affiliation{Korea Astronomy and Space Science Institute 776, Daedeokdae-ro, Yuseong-gu, Daejeon, 34055, Republic of Korea}
\affiliation{SOFIA Science Centre, USRA, NASA Ames Research Centre, N232 Moffett Field, CA 94035, USA}

\author{Paul F. Goldsmith}
\affiliation{Jet Propulsion Laboratory, California Institute of Technology, 4800 Oak Grove Drive, Pasadena, CA 91109, USA}

\author{Yu-Nung Su}
\affiliation{Academia Sinica, Institute of Astronomy and Astrophysics, P.O. Box 23-141, Taipei 106, Taiwan}

\author{Ken'ichi Tatematsu}
\affiliation{National Astronomical Observatory of Japan, 2-21-1 Osawa, Mitaka, Tokyo 181-8588, Japan}

\author{Sheng-Li Qin}
\affiliation{Department of Astronomy, Yunnan University, and Key Laboratory of Astroparticle Physics of Yunnan Province, Kunming, 650091, China}

\author{Guido Garay}
\affiliation{Departamento de Astronom\'{\i}a, Universidad de Chile, Casilla 36-D, Santiago, Chile}

\author{Tomoya Hirota}
\affiliation{National Astronomical Observatory of Japan, 2-21-1 Osawa, Mitaka, Tokyo 181-8588, Japan}

\author{Jan Wouterloot}
\affiliation{East Asian Observatory, 660 N. A'ohoku Place, Hilo, HI 96720, USA}

\author{Huei-Ru Chen}
\affiliation{Institute of Astronomy and Department of Physics, National Tsing Hua University, Hsinchu, Taiwan}

\author{Neal J. Evans II}
\affiliation{Department of Astronomy, The University of Texas at Austin, 2515 Speedway, Stop C1400, Austin, TX 78712-1205}
\affiliation{Korea Astronomy and Space Science Institute 776, Daedeokdae-ro, Yuseong-gu, Daejeon, 34055, Republic of Korea}

\author{Sarah Graves}
\affiliation{East Asian Observatory, 660 N. A'ohoku Place, Hilo, HI 96720, USA}

\author{Sung-ju Kang}
\affiliation{Korea Astronomy and Space Science Institute 776, Daedeokdae-ro, Yuseong-gu, Daejeon, 34055, Republic of Korea}

\author{Di Li}
\affiliation{National Astronomical Observatories, Chinese Academy of Science, A20 Datun Road, Chaoyang District, Beijing 100012, China}
\affiliation{Key Laboratory for Radio Astronomy, Chinese Academy of Sciences, Nanjing 210008, China}

\author{Diego Mardones}
\affiliation{Departamento de Astronom\'{\i}a, Universidad de Chile, Casilla 36-D, Santiago, Chile}

\author{Mark G. Rawlings}
\affiliation{East Asian Observatory, 660 N. A'ohoku Place, Hilo, HI 96720, USA}

\author{Zhiyuan Ren}
\affiliation{National Astronomical Observatories, Chinese Academy of Science, A20 Datun Road, Chaoyang District, Beijing 100012, China}

\author{Ke Wang}
\affiliation{European Southern Observatory, Karl-Schwarzschild-Str.2, D-85748 Garching bei M\"{u}nchen, Germany}

\begin{abstract}

How stellar feedback from high-mass stars (e.g., H{\sc ii} regions) influences the surrounding interstellar medium and regulates new star formation is still unclear. To address this question, we observed the G9.62+0.19 complex in 850 $\micron$ continuum with the JCMT/POL-2 polarimeter. An ordered magnetic field has been discovered in its youngest clump, the G9.62 clump. The magnetic field strength is determined to be $\sim$1 mG. Magnetic field plays a larger role than turbulence in supporting the clump. However, the G9.62 clump is still unstable against gravitational collapse even if thermal, turbulent, and magnetic field support are taken into account all together. The magnetic field segments in the outskirts of the G9.62 clump seem to point toward the clump center, resembling a dragged-in morphology, indicating that the clump is likely undergoing magnetically-regulated global collapse. However, The magnetic field in its central region is aligned with the shells of the photodissociation regions (PDRs) and is approximately parallel to the ionization (or shock) front, indicating that the magnetic field therein is likely compressed by the expanding H{\sc ii} regions that formed in the same complex.

\end{abstract}


\section{Introduction}

Stellar feedback from massive stars can exert a strong influence on the surrounding
medium and regulate the subsequent star formation \citep{el77,whit94a,whit94b}. The presence of feedback-influenced star formation process has been suggested in the borders of several H{\sc ii} regions as evidenced by fragmented shells, age sequence of stars or overdensity of young stellar objects \citep{za06,za07,thom12,liu12,liu15,liu16}. However, it is still unclear how stellar feedback from high-mass stars (e.g., H{\sc ii} regions) influences the surrounding interstellar medium and regulates new star formation.

Magneto-hydrodynamic (MHD) simulations demonstrated that the global magnetic field lines may roughly trace the outline of the expanding shell from a young massive star and are parallel to the long axis of the adjacent compressed filament \citep[e.g.,][]{klass17}, a strong evidence for stellar feedback. Recent near-infrared polarization observations have found that the magnetic field in the shells near H{\sc ii} regions or infrared bubbles is curved and following the shells, and the magnetic field strength in the shells is significantly enhanced compared to the ambient field strength \citep[e.g.,][]{chen17}. These near-infrared polarization observations are consistent with the simulations but only trace low-density, diffuse cloud material due to high dust extinction in the densest part of the shells.

In contrast to near-infrared polarization observations, polarized sub-millimeter thermal dust emission can trace magnetic field in dense regions of clouds \citep{hull13,zhang14}. To this end, we observed the G9.62+0.19 complex in 850 $\micron$ polarized continuum with the POL-2 polarimeter \citep{frib16} in conjunction with SCUBA-2 \citep{holland13} at the 15-m JCMT telescope. Located at a distance of 5.2 kpc \citep{san09}, the G9.62+0.19 complex is an active high-mass star forming region. Sequential high-mass star formation (from high-mass starless cores, hot molecular cores, UC~H{\sc ii} regions to expanding H{\sc ii} regions) is taking place in this region \citep{hof94,hof96,hof01,tes00,liu11,liu17}. Therefore, the G9.62+0.19 complex is an ideal target to study the effect of stellar feedback from expanding H{\sc ii} regions on next generations of high-mass star formation. \cite{liu17} suggested that the youngest star forming clump in this region (i.e., the G9.62 clump) is gravitationally unstable and will further collapse if only turbulent support is considered.

In this letter, we discuss the magnetic field geometry as well as magnetic field strength in the G.62 clump. In particular, we investigate how the magnetic field responds to stellar feedback and regulate the star formation in the G9.62 clump.

\section{Observations}

The POL-2 observations of the G9.62 clump (project code: M18BP019; PI: Tie Liu) were conducted in 2018 August using the POL-2 DAISY mapping mode \citep{holland13,frib16}. The total
integration time was 1.8 hr under JCMT Band 2 weather condition, with atmospheric optical depth at 225 GHz of $0.05<\tau_{225}<0.08$. The observing strategy is the same as described by \cite{ward17}. Data reduction is performed using a python script called \textit{pol2map} written within the STARLINK/SMURF package \citep{chap13,curr14}, which is specific for submillimetre data reduction (much of it specific to the JCMT). The output polarization percentage values are debiased using the mean of their Q and U variances to remove statistical biasing in regions of low signal-to-noise. The details of data reduction with \textit{pol2map} can be found in some previous POL-2 papers \citep{kwon18,liu18,pattle18,soam18}. Our method slightly differs from those previous works by utilizing the new \textit{skyloop}\footnote{http://starlink.eao.hawaii.edu/docs/sun258.htx/sun258ss72.html} parameter in \textit{pol2map} and the correction of synchronization loss between data values and pointing information in the data reduction process, which improve the ability to recover faint extended structures. The final co-added maps have rms noise levels of $\sim$4 mJy/beam for a beam size of 14.1$\arcsec$. Throughout this paper, polarization angles are measured from North increasing towards East, following the IAU convention. The polarization orientations obtained are rotated by 90$\arcdeg$ to infer the magnetic field orientations projected on the plane of sky.

Planck 850 $\micron$ (353 GHz) data are used to examine the dust emission and the dust polarization at scales larger than 5$\arcmin$ \citep{juvela18}.

\section{Results}

\begin{figure}[tbh!]
\centering
\includegraphics[angle=0,scale=0.4]{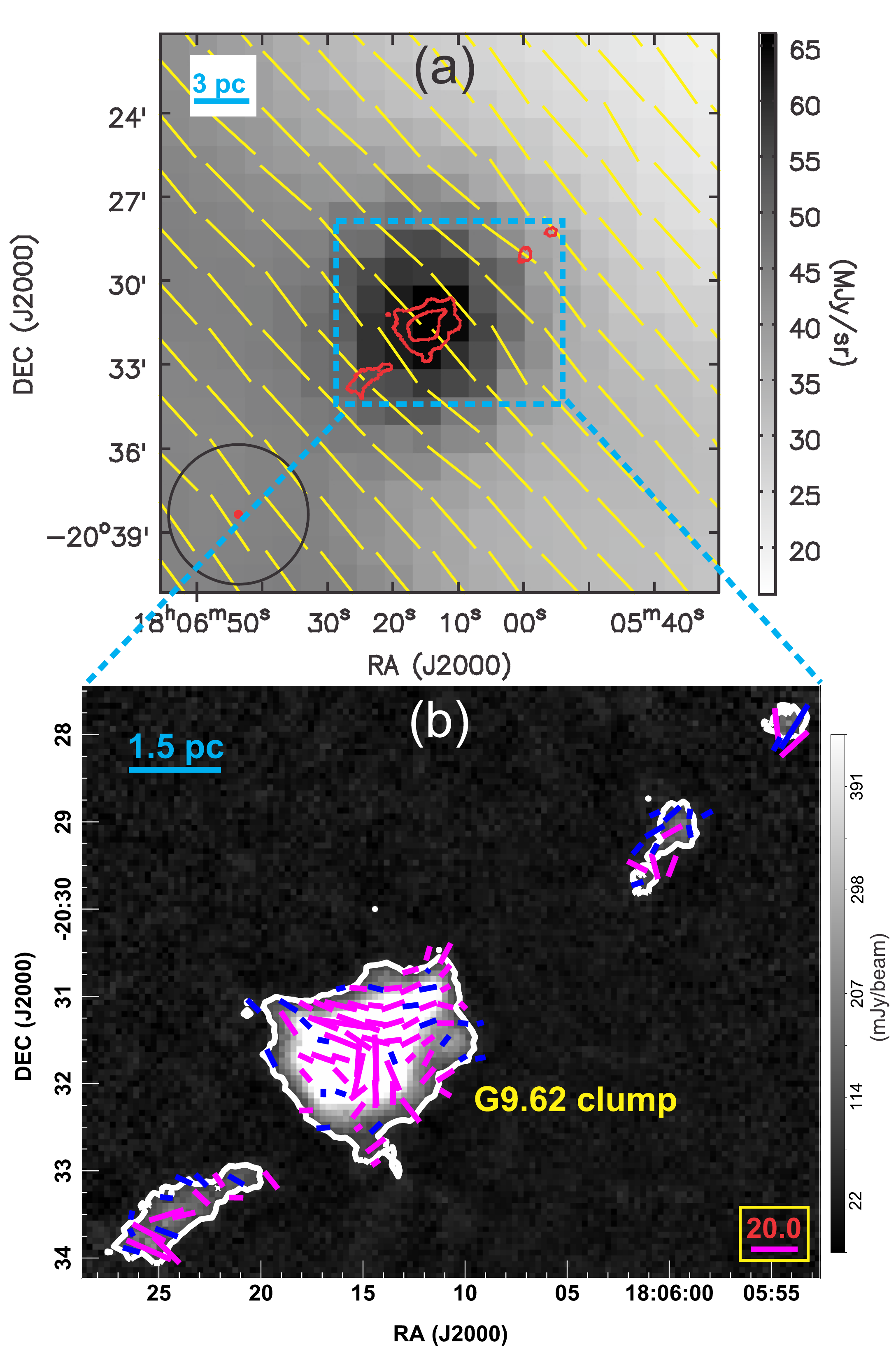}
\caption{(a) Planck 353 GHz magnetic field segments overlaid on the 353 GHz continuum emission shown in grey scale. The contours show the JCMT/POL-2 850 $\micron$ Stokes I intensity map. The contour levels are 50 and 500 mJy~beam$^{-1}$. The beams of Planck (black circle) and JCMT (filled red circle) are shown in the lower-left corner. (b) JCMT/POL-2 magnetic field segments overlaid on the Stokes I intensity map at 850 $\micron$. The contour is at 50 mJy~beam$^{-1}$. The segments with Stokes I intensity $I/\delta I>10$ are shown with a pixel size of 12$\arcsec$. The length of segments represents their polarized intensity in units of mJy~beam$^{-1}$.  The magenta segments have polarization fraction $P/\delta P>3$, while the blue segments have $2<P/\delta P<3$. \label{Bfields}}
\end{figure}

\begin{figure}[tbh!]
\centering
\includegraphics[angle=0,scale=0.4]{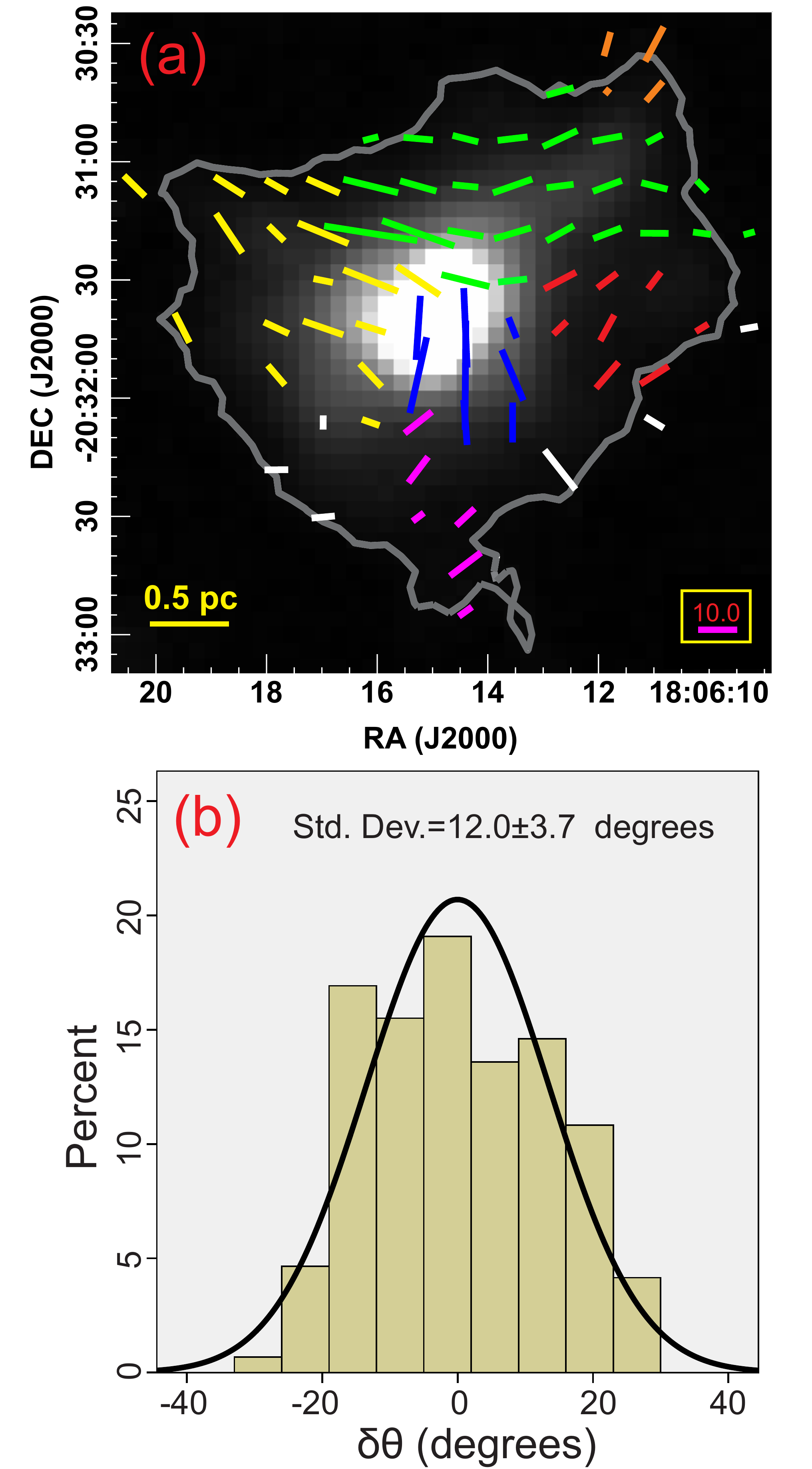}
\caption{(a) JCMT/POL-2 magnetic field segments overlaid on Stokes I intensity map at 850 $\micron$ in the G9.62 clump. The contour is at 50 mJy~beam$^{-1}$. The segments with polarization fraction $P/\delta P>2$ are shown with a pixel size of 12$\arcsec$. The different magnetic field components with underlying uniform field geometry are color-coded. The segments in white are not included in the analysis. (b) The distribution of the residual angles ($\delta\theta$) with $P/\delta P>2$. The distribution is weighted by $P/\delta P$. \label{dist}}
\end{figure}

\begin{deluxetable}{cccccc}[tbh!]
\centering
\tablecolumns{6} \tablewidth{0pc}\setlength{\tabcolsep}{0.05in}
\tablecaption{Statistics of the orientations of magnetic field segments \label{table}}\tablehead{\colhead{Component\tablenotemark{a}} & \colhead{Number} & \colhead{Min} & \colhead{Max}  & \colhead{Mean\tablenotemark{b}} & \colhead{$\sigma_{\theta}$\tablenotemark{b}}\\
\colhead{}  &\colhead{} & \colhead{($\arcdeg$)} & \colhead{($\arcdeg$)} & \colhead{($\arcdeg$)}  & \colhead{($\arcdeg$)} }
\startdata
Planck       &930 & 25.1 & 67.7  & 41.6 & 4.0 \\
JCMT-all     &110  & 2.4 & 180.0 & 91.1 & 44.8 \\
JCMT-red     &8   &117.7 & 151.5 & 130.7 & 11.0\\
JCMT-blue    &7   & -12.5\tablenotemark{c}& 23.2 & 1.3 & 11.4\\
JCMT-yellow  &17  & 26.8 & 78.7 & 58.2 & 13.1 \\
JCMT-green   &26  & 70.6 &119.1 & 93.0 & 16.1\\
JCMT-orange  &4   & 139.4&164.6 & 149.2 & 10.8 \\
JCMT-magenta &6  & 125.7 & 143.0 &130.8 & 6.4\\
\enddata
\tablenotetext{a}{Planck: Planck 353 GHz magnetic field segments within 30$\arcmin\times30\arcmin$ region; JCMT-all: JCMT/POL-2 magnetic field segments of the three clumps within the 50 mJy~beam$^{-1}$ contours of Stokes I intensity in panel (b) of Figure \ref{Bfields}; The ``JCMT-color" components correspond to the segments with different colors in panel (a) of Figure \ref{dist}. }
\tablenotetext{b}{The statistics of angles with $P/\delta P>2$. The mean angle and angle dispersion ($\sigma_{\theta}$) are weighted by $P/\delta P$.}
\tablenotetext{c}{The negative value is caused by the 180$\arcdeg$ ambiguity in magnetic field direction. A value of -180$\arcdeg$ is added to the segments with angles $>90\arcdeg$.}
\end{deluxetable}

\subsection{Orientations of magnetic field}

Panel (a) of Figure \ref{Bfields} shows the magnetic field morphology observed by Planck. The magnetic field orientations in the Planck data are quite uniform with a mean angle of $\sim42\arcdeg$ and a small angle dispersion of $\sim4\arcdeg$. The field direction is well aligned with the large scale ($\sim$100 pc) Galactic field direction. As shown in panel (b) of Figure \ref{Bfields}, four smaller clumps separated by $\sim$5 pc were detected in the POL-2 observations. Interestingly, the four clumps are aligned along a line that is roughly perpendicular to the large-scale magnetic field revealed by Planck, indicating that magnetic field may play an important role in the formation and fragmentation process of molecular clumps at the pc scale.

The magnetic field within the clumps as revealed by POL-2 shows a more complex behavior, with no preferred orientation. Although complexity is observed when considering the region as a whole, the field becomes more structured in the G9.62 clump as shown in Panel (a) of Figure \ref{dist}. We have identified six magnetic field components with underlying uniform field geometry. The neighbouring magnetic field segments with angle differences smaller than $\sim$15$\arcdeg$ from each other are assigned to the same component. These components are clearly separated from each other with the mean angles differing by $\gtrsim40\arcdeg$. They have small angle dispersions ($\lesssim16\arcdeg$), suggesting that the magnetic field is quite uniform within individual sub-regions of the G9.62 clump. They are color-coded in Panel (a) of Figure \ref{dist}. The statistics of those magnetic field components are shown in Table \ref{table}. The ``JCMT-yellow" component shows similar orientations ($\sim58\arcdeg$) as the large scale magnetic field revealed by Planck. The other components, however, show much larger deviation ($>50\arcdeg$) from the large scale magnetic field.

\subsection{Magnetic field strength}

We estimated the plane-of-sky magnetic field strength ($B_{pos}$) for the G9.62 clump using the Davis-Chandrasekhar-Fermi (DCF) method \citep{davis51,chandr53}: \begin{equation}
B_{pos}=Q'\sqrt{4\pi\rho}\frac{\sigma_{NT}}{\sigma_{\theta}}\approx9.3\sqrt{\frac{n(H_2)}{(cm^{-3})}}\frac{\Delta v/(km~s^{-1})}{\sigma_{\theta}/(degrees)}~\mu G,
\end{equation}
where $Q'$ is a factor of order unity accounting for variations in field strength on scales smaller than the beam \citep{crut04}, $\rho=\mu_g m_{H}n_{H_2}$ is the gas density. Here $Q'$ is taken as 0.5 \citep{ostr01}. We adopt the clump-averaged number density of $n_{H_2}=(9.1\pm0.7)\times10^4$ cm$^{-3}$ \citep{liu17}. $\Delta v$ is the FWHM velocity dispersion ($\sim$3.4$\pm$0.1 km~s$^{-1}$) derived from the C$^{18}$O (3-2) line observed with the JCMT \citep{liu17}. $\sigma_{\theta}$ is the dispersion in polarization position angles.

We subtract a mean angle from the measured position angles in each magnetic field component, giving residual angles ($\delta\theta$) showing the deviation in angle from the mean field direction. The distribution of the residual angles ($\delta\theta$) of the six magnetic field components is shown in panel (b) of Figure \ref{dist}. The dispersion in polarization position angle ($\sigma_{\theta}$) estimated from this distribution is $\sim13.4\pm3.7\arcdeg$. After correcting the angular dispersion for mean angle measurement uncertainty ($\sim6.1\arcdeg$), the $\sigma_{\theta}$ becomes $\sqrt{13.4^2-6.1^2}\approx12.0\arcdeg$. Because of significant changes in magnetic field orientations among different magnetic field components and the limited pixel numbers, we did not apply other methods (like the ``Unsharp Masking method" \citep{pattle17}) to remove the underlying uniform magnetic field. Without the exact knowledge of the uniform field, the derived $\sigma_{\theta}$ is not very precise. However, the $\sigma_{\theta}$, is smaller than the maximum value at which the standard DCF method can be safely applied \citep[$\leq25\arcdeg$; ][]{heit01}. Therefore, the derived $\sigma_{\theta}$ could be representative of the real angular dispersion, allowing us to perform an order-of-magnitude estimation of magnetic field strength and energies in the G9.62 clump.

The estimated $B_{pos}$ is $\sim790\pm190$ $\mu$G. Hereafter, the errors for derived parameters (e.g., $B_{pos}$) are estimated using standard error propagation. The magnetic field strength is significantly larger than that observed in some infrared dark clouds \citep[e.g., $\sim$270 $\mu$G in G11.11-0.12; $\sim$100 $\mu$G in G035.39-00.33; ][]{pillai15,liu18} on similar spatial scales. Statistically, the total magnetic field strength is 1.3 times $B_{pos}$ considering projection effects\citep{crut04}. Applying the same correction factor, the total magnetic field strength ($B_{tot}$) in the G9.62 clump is $\sim1030\pm250~\mu$G. We should note that the correction factor was derived from statistical studies and may not apply precisely to any individual region.

The corresponding Alf\'{v}enic velocity is:
\begin{equation}
\sigma_{A}=\frac{B_{tot}}{\sqrt{4\pi\rho}},
\end{equation}

The derived $\sigma_{A}$ is $\sim4.5\pm1.2$ km~s$^{-1}$. The Alf\'{v}en Mach number is:
\begin{equation}
\mathcal{M_A}=\sqrt{3}\sigma_{NT}/\sigma_{A}.
\end{equation}
where $\sigma_{NT}\sim1.5$ km~s$^{-1}$ is the mean non-thermal velocity dispersion derived from C$^{18}$O (3-2) line \citep{liu17}. $\mathcal{M_A}$ is $\sim0.6\pm0.2$, suggesting that the turbulent motions are sub-Alfv\'{e}nic in the G9.62 clump.

\section{Discussions}

\subsection{Gravitational stability of the G9.62 clump}

\cite{liu17} suggested that the G9.62 clump is gravitationally unstable if only turbulent support is considered. The question we wish to ask therefore is: does the magnetic field play a role in supporting the G9.62 clump?

To investigate the gravitational stability of the G9.62 clump, we estimated its virial mass ($M_{vir}^{B}$) considering thermal, turbulent, and magnetic pressure \citep{bert92,hen08,pillai11}:
\begin{equation}
M_{vir}^{B}=3\frac{R_{eff}}{G}(\frac{5-2n}{3-n})(\sigma_{NT}^2+C_{s}^2+\frac{\sigma_{A}^2}{6})
\end{equation}
where \textit{n} is power-law index for a density profile, $\rho(r)$, as a function of the distance (\textit{r}) from the clump center, $\rho(r)\propto r^{-n}$. \cite{muel02} derived a power-law index \textit{n}$\sim$2 for the G9.62 clump. $R_{eff}$ of 0.5 pc is the effective radius of the clump estimated from the SCUBA-2 850 $\micron$ continuum data \citep{liu17}. The 1D thermal velocity dispersion (or sound speed $C_s$) is 0.35 km~s$^{-1}$ for a temperature of 35 K \citep{liu17}. For a $\sigma_{A}$ of $\sim$4.5 km~s$^{-1}$ and a $\sigma_{NT}$ of $\sim$1.5 km~s$^{-1}$, the virial mass is $\sim1900\pm600$ M$_{\sun}$, which is smaller than the clump mass $\sim$2800$\pm$200 M$_{\sun}$ \citep{liu17}, indicating that the G9.62 clump is unstable against gravitational collapse even if thermal, turbulent, and magnetic field support are taken into account all together. Magnetic field plays a ($\sim$1.5 times) larger role than turbulence in supporting the clump.

We also notice that most of the magnetic field segments (like the white, yellow, and orange segments in panel (a) of Figure \ref{dist}) in the outskirts of the G9.62 clump seem to point toward the clump center, resembling a dragged-in morphology as seen in other magnetically-regulated collapsing cores \citep[e.g.,][]{tang09,koch18}. The magnetic field geometry and the gravitational stability of the G9.62 clump indicate that it may be undergoing magnetically-regulated global collapse.

\subsection{Compressed magnetic field due to stellar feedback from expanding H{\sc ii} regions}

\begin{figure*}[tbh!]
\centering
\includegraphics[angle=0,scale=0.5]{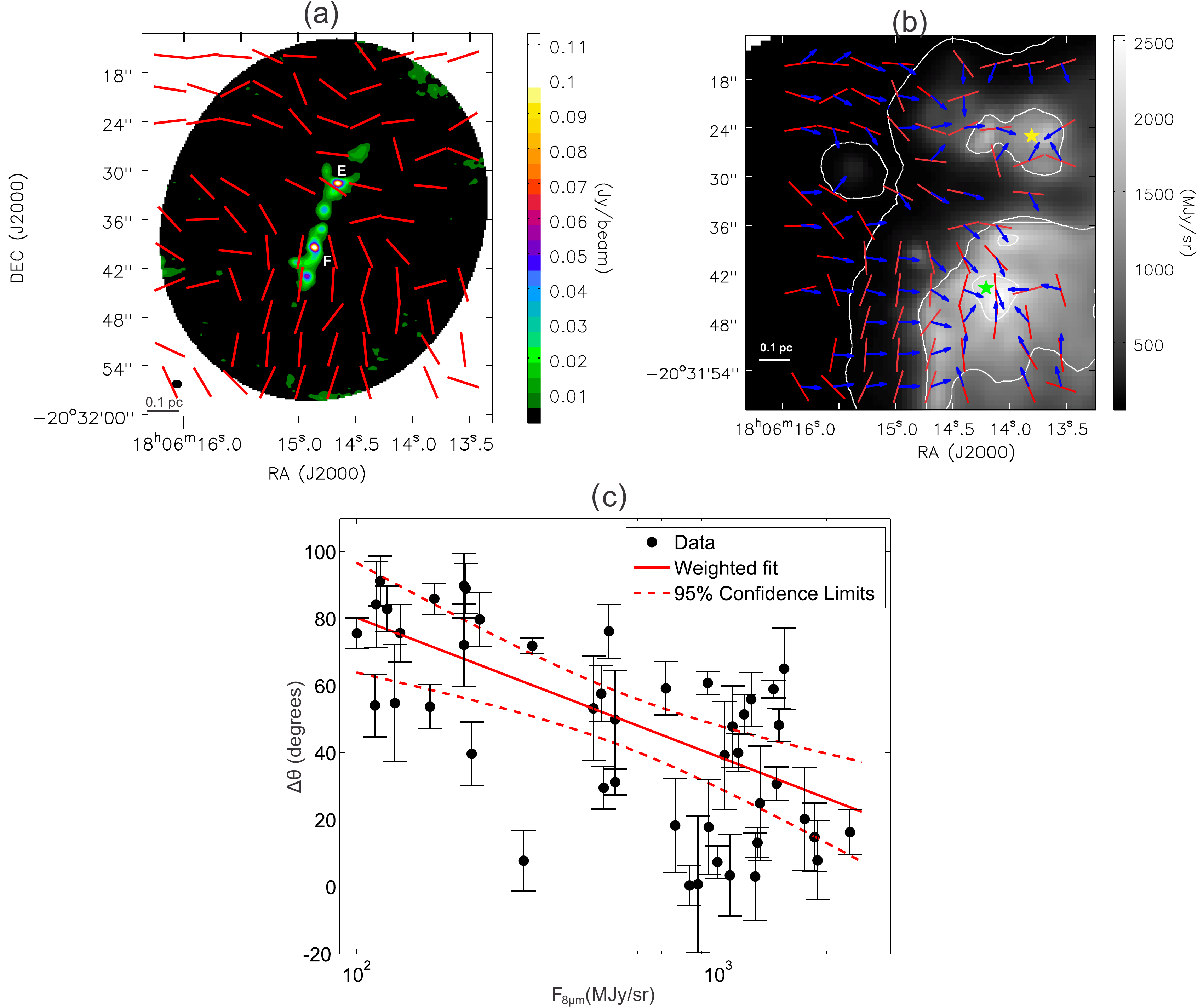}
\caption{(a) JCMT/POL-2 magnetic field segments overlaid on the ALMA 1.3 mm continuum image \citep{liu17}. The pixel size of POL-2 data shown here is 4$\arcsec$. The ALMA beam is shown as the filled circle in the lower-left corner. The two brightest cores in the ALMA image are ``E" (northern one) and ``F" (southern one). (b) JCMT/POL-2 magnetic field segments (red) overlaid the Spitzer/IRAC 8 $\micron$ emission shown as grey image and white contours. The contour levels are 100, 1000 and 2000 MJy/sr.  The two expanding H{\sc ii} regions are marked with stars (``B" in green and ``C" in yellow). The blue arrows show the directions of Spitzer/IRAC 8 $\mu$m intensity gradients.  (c) The angle differences ($\Delta\theta$) between magnetic field segments and 8 $\mu$m intensity gradients as a function of 8 $\mu$m intensity (F$_{8\mu m}$). The solid red line is the best fit to the data considering the uncertainties of magnetic field angles. The dashed red lines are the confidence ranges of the best fit. \label{comp}}
\end{figure*}

Although the G9.62 clump seems to be undergoing magnetically-regulated global collapse, its magnetic field morphology also indicates influences of stellar feedback from the older generations of expanding H{\sc ii} regions formed in the same complex.

As shown in panel (a) of Figure \ref{dist}, the magnetic field segments (e.g., JCMT-blue and JCMT-green components) in the central region do not follow a dragged-in (e.g. hour-glass) morphology caused by gravitational collapse. In the panel (a) of Figure \ref{comp}, we take a closer look at the magnetic field segments in its central region. The dense filament as revealed by the ALMA 1.3 mm continuum (shown as color image in the panel (a) of Figure \ref{comp}) has a position angle of $\sim162\arcdeg$. The magnetic field segments are roughly parallel to the dense filament in its southern part, while the magnetic field segments become roughly perpendicular to the dense filament in its northern part. Previous polarization observations of filamentary clouds \citep{chap11,cox16,liu18,juvela18} found that magnetic field tends to be roughly perpendicular to the longer axes of the dense parts of filaments, indicating that the dense filaments are collapsing along magnetic field or still accreting gas along magnetic field from their surroundings, which is very different from the dense filament in the G9.62 clump. \cite{liu17} suggested that the G9.62 clump is compressed by the expanding H{\sc ii} regions (``B" and ``C") to its west. Therefore, below we argue that the magnetic field in the G9.62 clump seems to be compressed as the expanding H{\sc ii} regions grow.

The Spitzer/IRAC 8 $\micron$ emission is mainly dominated by polycyclic aromatic hydrocarbons (PAH) emission and is a good tracer of PDRs \citep{chur06}. The 8 $\micron$ emission in the panel (b) of Figure \ref{comp} reveals the PDRs of the two expanding H{\sc ii} regions. We gridded the 8 $\mu$m data to a 4 arcsec pixel size, the same as POL-2 data, to derive the 8 $\micron$ intensity gradients with the function ``gradient"\footnote{https://www.mathworks.com/help/matlab/ref/gradient.html} in MATLAB. As shown in the panel (b) of Figure \ref{comp}, the magnetic field segments in the outskirts of ``B" and ``C" roughly follow the 8 $\micron$ intensity contours and are perpendicular to the 8 $\mu$m intensity gradients. In the panel (c) of Figure \ref{comp}, we investigate the angle differences ($\Delta\theta$) between magnetic field segments and 8 $\mu$m intensity gradients as a function of 8 $\mu$m intensity (F$_{8\mu m}$) for pixels with F$_{8\mu m}>$100 MJy/sr. A clear decreasing trend in the $\Delta\theta$ vs. F$_{8\mu m}$ relation is found as:
\begin{equation}
\frac{\Delta\theta}{(degrees)}=-(18.0\pm3.3)\textrm{ln}(\frac{F_{8\mu m}}{(MJy/sr)})+(163.3\pm21.2),
\end{equation}
with the correlation coefficient R=0.63; further indicating that the magnetic field surrounding the expanding H{\sc ii} regions becomes to follow the outlines of the expanding shells and is approximately parallel to the ionization (or shock) front. In numerical simulations of expanding H{\sc ii} regions, a shell of material is swept up as the H{\sc ii} region grows and the magnetic field inside the shell is approximately parallel to the ionization front \citep{arth11,klass17}, which is consistent with our findings here.

In numerical simulations, the magnetic field strength is enhanced by a factor of about 5 to 6 in the compressed shell when comparing the magnetic field strength inside the expanding H{\sc ii} regions \citep{klass17}. If we adopt the same enhancement factor, the magnetic field strength inside the expanding H{\sc ii} regions (``B" and ``C") should be $\sim$200 $\mu$G. The total magnetic pressure ($P_{B}$) is:
\begin{equation}
P_{B}=\frac{B_{tot}^2}{8\pi k_B}
\end{equation}
where $k_B$ is the the Boltzmann constant. The estimated $P_{B}$ inside the H{\sc ii} regions is $\sim1\times10^7$ K~cm$^{-3}$, which is smaller than the ionized gas pressure ($\sim4\times10^7$ K~cm$^{-3}$) derived by \cite{liu17}, suggesting that the magnetic field cannot prevent H{\sc ii} regions from further expanding. The magnetic field may also have being compressed as the H{\sc ii} regions expand \citep{arth11,klass17}.

\subsection{Comparison with core-scale magnetic field}

\cite{dall17} recently observed the magnetic field at $\sim$336 GHz around dense cores ``E" and ``F" (the two brightest cores in panel (a) of Figure \ref{comp}) with the ALMA. The magnetic field around ``E" roughly follows an east-west direction and is perpendicular to the filament long axis, while the magnetic field around ``F" is parallel to the filament. The orientations of the magnetic field segments in the ALMA observations are consistent with those in our JCMT/POL-2 observations. This consistency of clump-scale and core-scale magnetic field orientations may suggest that the fragmentation in the G9.62 clump is regulated by magnetic field.

\cite{liu17} found a lack of a widespread low-mass protostellar population and suggested that the core fragmentation or low-mass star formation is suppressed due to feedback from young OB stars in the G9.62 clump by heating the cores up and injecting turbulence through outflows, leading to an increase of the Jeans mass. MHD simulations suggested that the combination of magnetic field and radiation feedback is even more effective at suppressing fragmentation \citep{myers13}. The very ordered magnetic field revealed by the ALMA around the two bright cores ``E" and ``F"  \citep{dall17} indicates strong magnetic field strength at the core scale. Therefore, we suggest that core fragmentation in the G9.62 clump is very likely suppressed due to the joint effect of the strong magnetic field and feedback (e.g., radiation, outflows) from young OB stars. Indeed, higher angular resolution (0.3$\arcsec$ or $\sim$1500 AU) ALMA observations at 850 $\micron$ indicate that the massive cores in the G9.62 clump are not highly fragmented \citep{dall17}.

\section*{Acknowledgment}
\begin{acknowledgements}

The James Clerk Maxwell Telescope is operated by the East Asian Observatory on behalf of The National Astronomical Observatory of Japan; Academia Sinica Institute of Astronomy and Astrophysics; the Korea Astronomy and Space Science Institute; the Operation, Maintenance and Upgrading Fund for Astronomical Telescopes and Facility Instruments, budgeted from the Ministry of Finance (MOF) of China and administrated by the Chinese Academy of Sciences (CAS), as well as the National Key R\&D Program of China (No. 2017YFA0402700). Additional funding support is provided by the Science and Technology Facilities Council of the United Kingdom and participating universities in the United Kingdom and Canada. The Starlink software \citep{curr14} used in POL-2 data reduction is currently supported by the East Asian Observatory. Tie Liu is supported by EACOA fellowship. M.J. acknowledges the support of the Academy of Finland grant no. 285769. This work was carried out in part at the Jet Propulsion Laboratory, which is operated for NASA by the California Institute of Technology. Yuefang Wu acknowledges the grants of the National Key R\&D Program of China No. 2017YFA0402600 and NSFC Nos. 11433008.

\end{acknowledgements}

\end{document}